%% file: ME667rv2_title.tex
\documentstyle[graphicx,psfig,amssymb,longtable]{mn2e}

\newcommand{\hMpc}{{\ifmmode{h^{-1}{\rm Mpc}}\else{$h^{-1}$Mpc}\fi}}
\newcommand{\hkpc}{{\ifmmode{h^{-1}{\rm kpc}}\else{$h^{-1}$kpc}\fi}}

\def\approxlt{\mathrel{\spose{\lower 3pt\hbox{$\sim$}}
        \raise 2.0pt\hbox{$<$}}}
\def\approxgt{\mathrel{\spose{\lower 3pt\hbox{$\sim$}}
        \raise 2.0pt\hbox{$>$}}}
\def\approxpropto{\mathrel{\spose{\lower 3pt\hbox{$\sim$}}
        \raise 2.0pt\hbox{$\propto$}}}

\input my_def_aa.txt
\def \beta2 { {b} }

 \title{       Period recovery success rate expected with Gaia}
\author[L. Eyer \& F. Mignard]{
                    L. Eyer$^{1,2}$ \& F. Mignard$^{3}$\\
                $^{1}$ Geneva Observatory, CH-1290 Sauverny \\
                $^{2}$ Princeton University Observatory, Princeton, NJ 08544, USA\\
                $^{3}$ Observatoire de la c\^ote d'azur, CNRS, BP 4229 , 06304,  Nice, France
          }
\pagerange{\pageref{firstpage}--\pageref{lastpage}}
\pubyear{2004}

\begin{document}

\maketitle

\label{firstpage}

  \begin{abstract} The \gaia\ satellite was selected as a cornerstone mission
    of the European Space Agency (ESA) in October 2000 and confirmed
    in 2002 with a current target launch date of 2011. The
    \gaia\ mission will gather on the same observational
    principles as \hip\ detailed astrometric, photometric and
    spectroscopic properties of about 1~billion sources brighter than
    mag $V=20$. The nature of the measured objects ranges from NEOs
    to gamma ray burst afterglows and encompasses virtually any kind
    of stars in our Galaxy. \gaia\ will provide multi-colour (in about
    20 passbands extending over the visible range) photometry with
    typically 250 observations distributed over 40 well separated epochs
    during the 5-year mission. The multi-epoch nature of the project
    will permit to detect and analyse variable sources whose number is
    currently estimated in the range of several tens of million, among the
    detectable sources.

    In this paper, we assess the performances of \gaia\ in analysing
    photometric periodic phenomena. We first present quickly the
    overall observational principle before discussing the implication
    of the scanning law in the time sampling. Then from extensive
    simulations one assesses the performances in  the recovery of
    periodic signals as a function of the period,  signal-to-noise
    ratio and  position on the sky for simple sinusoidal variability.
\end{abstract}


\section{Introduction: The \gaia\ mission }

The very successful ESA mission {\sc Hipparcos} was the first modern
multi-epoch all sky space survey. Although primarily designed for
astrometry the full light (supplemented by two chromatic bands of
lesser accuracy) photometry proved to be a significant addition to the
initial goal. About 10\% of the objects in the final catalogue have
information about variability, 7\% are listed in the periodic and
unsolved catalogues, and 3\% (i.e 3794) have been published with their
light curves. A refined analysis of the occurrence of variability
through the whole HR diagram was subsequently achieved
\cite{eyer97}.

The \gaia\ mission, a cornerstone of the ESA science programme,
will dramatically improve this analysis in several respects: (i)
the photometric precision will be much better, in the range of
milli-magnitude for single observations (compared to 0.01 for
\hip); (ii) the number of objects will be increased by a factor
nearly 10\,000; (iii) multiband photometry (15 to 20 photometric
channels in addition to the full light) will be available and will
permit to identify the physical processes responsible for the
variable phenomena. In addition, there will be an instrument for
low resolution spectroscopy. The mission selected in 2000 and
confirmed in 2002 with a modified design is scheduled for a launch
in mid-2011 for a 5-years sky survey.

Scientific preparation for the mission involves the participation
of some 15 working groups sharing responsibility for the
simulation, the data processing, the science modelling and the
instrument optimisation.  Extensive simulation programs have been
developed feeding all the groups with input data required to test
the processing. In particular the scanning parameters are now
fixed and constitute the basic tool to determine the times series.
The algorithm used in this paper is based on the analytical
algorithm built and implemented by one of us \cite{mignard01} and
used thoroughly by the Gaia community.

Regarding the photometry, as of mid-2005, the choice of passbands,
and the parameters of the  scanning law are now fixed. The
relative positions of the fields of view may again change before
the design is frozen by the prime contractor next year. We have
used the design of the study phase of mid-2003. The measurement
errors have been also assessed and for the full light magnitude
(the so-called $G$ magnitude) can be found in \cite{jordi03}. It
is better than 0.002 mag for a single observation (one transit in
the astrometric field) of stars brighter than $G = 15$ and of the
order of 0.01 mag for the faintest stars ($ G\sim 20$) detectable
with \gaia. This gives already a rough idea of the wealth of
science information recoverable on the variable stars, virtually
over any spectral type or luminosity class.

The large scale photometric survey will have significant value for
stellar astrophysics, supplying basic stellar parameters (effective
temperatures, metallicity and abundances, gravities) together with a
virtually unbiased sample of variable stars of nearly all types, such
as eclipsing binaries, pulsating stars with periods ranging from hours
to several hundred days. Much shorter periods could be achieved also
from the exploitation of the light recorded during the passage of the
image on individual CCDs of a field of view. This particular and very
specific sampling is not addressed in this paper.  Variability search
methods will be systematically applied to every photometric time
series, before the period search. Estimated number of variable stars
to be detected as such is highly uncertain, but crude estimates
suggest some 20 million in total, half of them being periodic
variables \cite{eyer00}.

\section{The observing instrument}

Although most technical details can be found on the \gaia\
documentation (e.g Perryman et al. 2001), there are two items
particularly relevant to the study of the variable stars, as they
impact directly on the time sampling. The first is the way the sky is
scanned, allowing to observe a given region of the sky at about 30 to
50 different times during the 5-year mission, with an average return
in the same direction every 6 weeks. The second is concerned with the
repeated measurements over much smaller time intervals, typically
between one to ten hours. This sampling is determined by the relative
positions of the different fields of view, through which photometric
measurements are carried out. These two aspects of the design are
discussed in the following.

The instrumental parameters used in this paper refer to the current
baseline of the design at the time of writing. This is subject to
possible change during the study phase, without major impact on the
main results of this paper, regarding the determination of the period
of the variables, all the more because the simulation is primarily
based on the data coming from the astrometric fields of view.

\subsection{The scanning law}

Measurements with \gaia\ are conducted with a scanning satellite with
two widely separated astrometric fields of view, similar in its
principle to the solution adopted for Hipparcos. This proved to be a
very efficient and reliable solution for global astrometry.

\begin{figure}
  \centerline{\includegraphics[width=7cm]{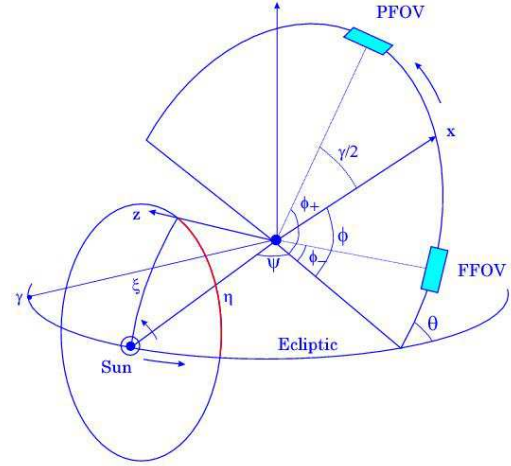}}
  \caption{The nominal scanning law of \gaia. The spacecraft rotates
    around the $z-$axis in 6 hours so that the \fov s sweep
    approximately a circle on the sky. The $z-$axis is constrained to
    move on a sun-centered cone of 50 degrees half-aperture with a
    period of 70 days, forcing the plane of scan to sway back and
    forth with inclinations between 40 to 140 deg. The axis of the
    cone follows the yearly solar motion.}\label{scan}
\end{figure}

The sky is scanned continuously following a pre-defined pattern, with
the satellite spin-axis kept at a fixed angle of 50\degr\ from the Sun
(Fig.~\ref{scan}) and nominally perpendicular to each viewing
direction. A sun-centered precession motion of the spin axis in $\sim
70$ days allows a rather uniform coverage of the full sky every six
months, yielding repeated observations of every region of the sky. The
scan period is six hours amounting to a displacement rate of the
stellar images on the \ccds\ of 60~arcsec/s.

\subsection{The photometric fields of view}

The \gaia\ payload is fitted with two main telescopes with an aperture
of $1.4\times 0.5$~m$^2$, primarily dedicated to astrometry and full
light photometry. On each field, several columns of \ccds\ are covered
with broad-band filters. This broad-band photometric field provides
multi-colour, multi-epoch photometric measurements for each object
observed in the astrometric field, for chromatic correction and
astrophysical purposes. Four or five photometric bands should be
implemented within each instrument with an integration time of $\sim
3.3$~s per \ccd. A third telescope with a square entrance pupil of
$0.5\times 0.5$~m$^2$ collects starlight for the spectrometer and the
medium-band photometer.

\section{Sequences of observation}

Due to the combination of the quick spin motion over six hours and
that of precession over 70 days coupled with the annual solar
motion, the sampling of the epochs at which a particular object is
measured is very peculiar and quite irregular. However, irregular
does not mean random, and there are some basic recurring patterns
which apply to each sequence. One must consider separately the
succession of observations over period less than $\sim$ 1-2 day
due to the repeated passages over few consecutive rotations of the
satellite and are primarily governed by the spin and the focal
plane layout, and the return in a particular direction after
several weeks without observations, controlled by the precession
motion. So a complete sequence divides itself naturally in
succession of epochs, widely separated in time, in which there are
a small number (between $\sim$ 1 to 10) of observations in the
astrometric or photometric fields of view. For a regular sampling
the ability to carry out accurate frequency analyses of variable
stars depends on the succession of repeated observations over
short time-scale, that is to say on sequences of successive
revolutions separated by returns of the viewing direction to the
same area of the sky. However for a purely random sampling one can
obtain alias-free results even for frequencies much higher than
the inverse of the smallest intervals between observations and one
should not worry too much about the time gaps between successive
observations. But Gaia sampling is obviously not perfectly regular
but also falls  short from a purely random sampling. This is
something intermediate with regular patches over short time-scales
repeating with much irregularity over longer time-scales. At the
end the combination of the two interleaved sampling proved
important to achieve an ambiguous and accurate frequency
retrieval.

The variety of short term intervals likely to occur within a sequence
of observations is given in Table~\ref{tab:interval}. The highest
temporal resolution is obtained by combining successive crossings in
the MBP field with a photometric measurement obtained in the
astrometric preceding field, leading to a minimum interval of 38
minutes.

\begin{table}
\setlength{\tabcolsep}{2.5mm}
\begin{center}
  \caption{Fundamental sampling intervals following from the \gaia\
    scanning law and from the location of the viewing directions.
    Transitions from an initial field to a final field are considered
    with the interval given in days. PFOV: preceding astrometric field
    of view, FFOV: following astrometric field of view, MBP: Medium
    band photometer field of view. }\vspace{5mm}
  \begin{tabular}{llr}
   \hline \\[-4pt]
      \cent{Initial field} &   \cent{Final field} &   \cent{Interval}  \\[2mm]
      && \cent{hours} \\[3mm]
    \hline\\[-4pt]
 PFOV    & FFOV     & 1.78  \\[1mm]
 PFOV    & MBP      & 5.38  \\[1mm]
 PFOV    & PFOV     & 6.00   \\[1mm]
 FVOV    & MBP      & 3.60   \\[1mm]
 FFOV    & PFOV     & 4.22  \\[1mm]
 FFOV    & FFOV     & 6.00   \\[1mm]
 MBP     & PFOV     & 0.62  \\[1mm]
 MBP     & FFOV     & 2.40   \\[1mm]
 MBP     & MBP      & 6.00   \\[2mm]
  \hline
  \end{tabular}
\label{tab:interval}
\end{center}
\end{table}

\subsection{Short term succession}
\subsubsection{Astrometric fields}

Let a star be observed in the preceding astrometric field, at some
reference time $t=0$.  The star is located somewhere in the field of
view of width $0.66$ degree in the cross-scan direction. We use a
local rectangular frame with x-axis along the scan direction and
y-axis in the transverse direction. Take the star image at the middle
of the field for this reference time, that is to say $y=0$. Six hours
later this viewing direction is back in nearly the same region of the
sky. However the pole of the scan circle has moved by nearly 1~degree
on the sky, and the ordinate of the star in the field can now be as
large as 1 degree. In this case the star is no longer in the field of
view and cannot be re-observed. On the other hand if the star happens
to lie very close to one of the two stationary points of the scan
circle (over a period of few days the scan of the sky can be described
by the slow motion of a great circle, implying that there are two
nodes on this circle which are stationary) the star has virtually not
moved on the field and will be observed again after one revolution,
and this situation could recur during four or five rotations of the
satellite, supplying as many observations in the preceding \fov.

Consider now the transition between the preceding and following
astrometric \fov's. They are separated by 106 degrees, meaning that it
takes 106 minutes to rotate from one field to the next.  During this
interval the apparent transverse motion of the star with respect to
the scan circle is at most of 0.3 degrees.  Therefore a star observed
at a random position in the preceding field will be observed in the
following field with a $\sim$ 0.5 probability. The average transverse
motion being only $2/\pi\times 0.3 = 0.19$~degree, this makes the odds
more favourable for the observations to be grouped at least by pair.
With smaller transverse motion (star closer to the instantaneous node)
one may have four or five successive pairs of observations, each
separated by 0.25 day. During these crossings there are photometric
measurements in full light and in the five broad band filters as well.

We can do a simple probabilistic modelling to assess the
probability that an observation in a particular field is isolated
(sequence of length 1) or followed by at least one observation in
the next available field of view. It is reasonable to assume that
the first observation of a sequence may occur at any ordinate in
the field (this is a crude approximation, as the first observation
of a sequence is more likely to occur near the upper or lower edge
of the field). The chance of having the star observable in the
next field depends on the transverse velocity and on the interval
of time between the two fields. Let $V_t$ be the transverse
velocity and $\Delta t$ the interval of time. One has $\Delta t =
106$ minutes from preceding \fov\ (\pfov) to following \fov\
(\ffov), 254 mn from \ffov\ to \pfov\ and 360 mn = 0.25 day for
one revolution. The transverse velocity depends very simply on the
longitude $\phi$ of the optical axis  on the instantaneous scan
circle and takes the form $ V_t = V_{\mbox{max}}\,\sin(\phi +
\alpha)$ with $V_{\mbox{max}} = 0.170$~deg/h and where $\phi$ is a
fast moving angle (period of 6 h) and the phase $\alpha$ is a slow
varying variable with a typical time-scale of 70 days. We call
$\mathbf{X}$ the random variable $\mathbf{X}=1$ if the star is not
re-observed after the first crossing at the start of a new epoch,
and $\mathbf{X}=0$ for the opposite outcome. If $h$ is the
transverse size of the field, it is easy to establish (see the
Appendix) by conditioning the random variable to the transverse
velocity and computing the expectation $
\mbox{E}(\mbox{E}(\mathbf{X}|\mathbf{V_t}))$ that,

\[\label{proba}
  P( \mathbf{X}=1) =
\]
\begin{equation}\label{probability}
      \left\{\begin{array}{ccl}
    \frac{2}{\pi}\, \frac{1}{u} & \mbox{\ if \ } & u \ge 1 \\[4mm]
    \frac{2}{\pi}\, \frac{1}{u}\,\left(1-(1-u^2)^{1/2}\right) + \frac{\pi -2\,\arcsin (u)}{\pi}& \mbox{\ if \ } & u \le 1
  \end{array}
  \right.
\end{equation}

where $ u = h/V_{\mbox{max}}\, \Delta t$. When $u > 1$ the
transverse motion over $\Delta t$ is always less than the width of
the \fov\ while it can be larger otherwise when $u < 1$. For the
transitions between the preceding to following \fov\ this yields a
probability $P = 0.71$ that more than one observation occurs and
0.32 for the transition between the following and preceding field,
with the same assumption that the observation in the first field
is the first of a sequence.

Table~\ref{tab:sequences} gives the actual frequency of the
different length of the sequence based on a simulation over
10\,000 randomly distributed stars on the sky. The positions of
these stars have been compared with the viewing directions over
five years, thus producing for each star the time sampling in each
field of view.  The termination criterion for a continuous
sequence of observations is activated when the interval between
two successive observations is larger than few revolution periods
of the satellite. A gap of 3 or 4 revolutions ($\sim 1$ day) may
happen in a short time sequence in the passage from MBP transits
to BBP transits as a result of the relative location of their
respective detectors as shown in Fig.~\ref{size}. From trials and
errors we have found that a time gap between two successive
observations larger than 2.5 days was a good signature to isolate
short observing sequences from very different observing epochs.
This means essentially that when one has a gap of 2.5 days without
observation, the actual gap will be in practice much larger ($\sim
20-30$ days), with few exceptions well accounted for by the
idiosyncrasies of the scanning law.

 In the first two columns of
Table~\ref{tab:sequences}, only the observations in the
astrometric fields have been monitored and sequences starting with
the preceding \fov\ (sequences like P, PF, PFP, PFPF $\cdots$) or
following \fov\ (sequences like F, FP, FPF, FPFP, $\cdots$) are
counted separately. They are not symmetrical, because the
occurrence of a sequence \pfov\ $\rightarrow$ \ffov\ is more
likely than the opposite sequence.  The typical sequence comprises
two observations (PF) in the first group, while sequences of unit
length (F) are the most common in the second group. The average
length of the sequences is 2.1 in the first case and 1.6 in the
second. While the number of long sequences is small (less than one
percent of the sequences longer than 10 observations), the length
of the sequences can be very large and can reach 20 or 40
consecutive observations, meaning that the star is observed at
regular interval for up to 5 days.

\subsubsection{Astrometric and spectroscopic fields}
\begin{figure}
  \centerline{\includegraphics[width=8cm,clip=true]{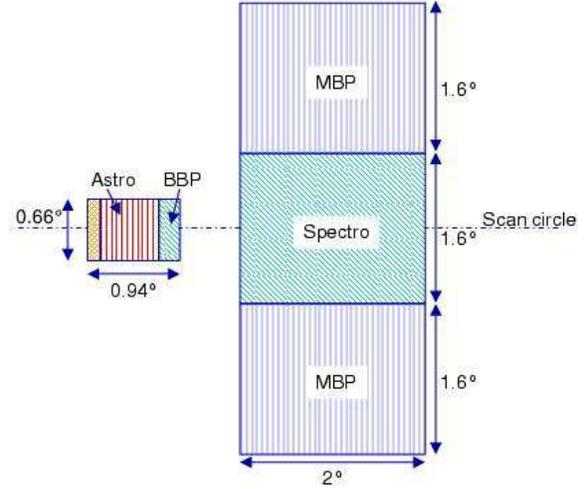}}
  \caption{Relative size of the fields of view of \gaia. Many more
    field crossings arise in the medium band photometer ({\sc mbp})
    than in the broad band photometers ({\sc bbp}) in the astrometric
    fields, as a result of the vertical extension. }\label{size}
\end{figure}

\begin{figure}
  \centerline{\includegraphics[width=8cm,clip=true]{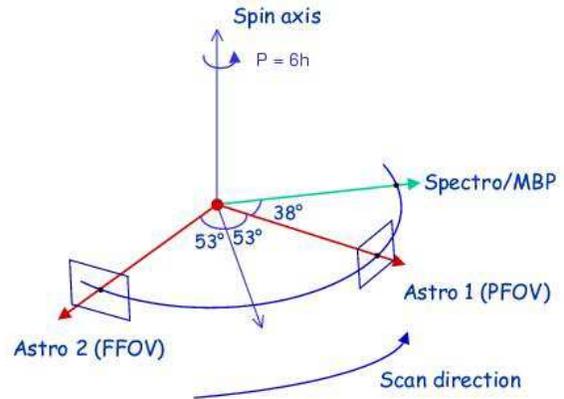}}
  \caption{Relative positions of the viewing directions along the
    scan circle. MBP = Medium band photometry, PFOV (resp. FFOV) =
    preceding (resp. following) field of view.}\label{fovs}
\end{figure}

\begin{figure}
  \centering%
  \mbox{\psfig{file=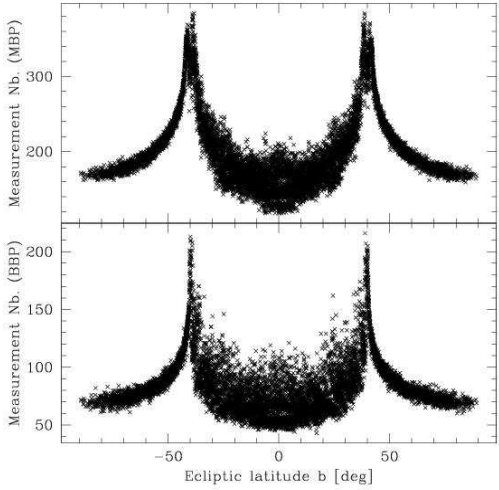,height=70mm,width=70mm}}%
  \caption{\label{fig:nmesbeta}
    Number of measurements in MBP (top) and BBP (bottom) fields of view
    as function of ecliptic latitude $ \beta2 $.  The mean number of
    observations is 200 for the MBP and 80 for the BBP. }
\end{figure}

We introduce now the field with the medium band filters (MBP fields of
view). It is located ahead of the preceding field, at 38 degrees as
depicted in Fig.~\ref{fovs} and comprises two subfields symmetrically
placed with respect to the scan direction as shown in Fig.~\ref{size}.
Its vertical extension is much larger than the astrometric fields,
meaning that the probability that a star is observed in this
photometric field is higher than in the astrometric \fov s.  The
average number of observations is just above 200 in the medium band
photometer over the 5-year mission.  The distribution with the
ecliptic latitude (Fig.~\ref{fig:nmesbeta}) displays the same
symmetrical behaviour as for the observations with the broad band
filters.

The transverse motion over one revolution is always smaller than
the vertical extension of any of the two subfields, which implies
that observations take place over at least two successive
revolutions of the satellite. In fact, the average length of a
sequence consists of 6-7 consecutive observations within less than
a day, with measurements in every bandpass. In this case all the
sequences but a handful start with an observation in the MBP field
and a typical sequence is MMPFMM, meaning two MBP measurements
followed by observations in the preceding and following
astrometric fields and then again two in the MBP field. The
probability distribution of the sequence lengths is given in the
rightmost column of Table~\ref{tab:sequences}. Very long sequences
of nearly 100 uninterrupted consecutive observations over $\sim$
10 days may happen exceptionally.

\begin{figure}
\centerline{\includegraphics[width=7cm]{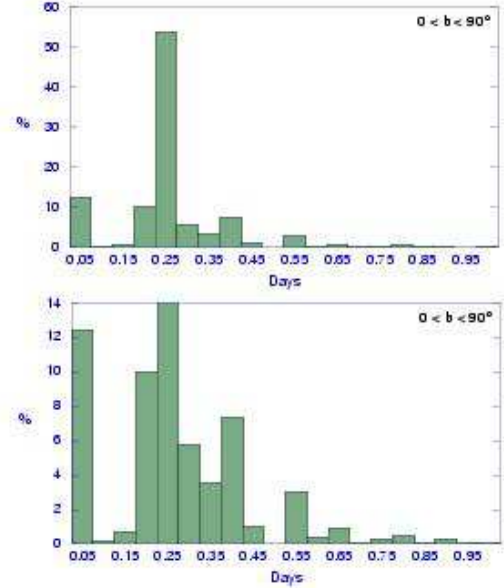}}
\caption{Frequency distribution of the intervals between
  consecutive observations performed at the same epoch in the three
  photometric fields of view. This indicates that the most common
  interval between paired observations is the 6 hours of the scan
  period.  The lower panel is a close-up view of the overall
  distribution shown in the upper panel. }\label{shortp}
\end{figure}

The distribution of the short intervals when observations in the
three photometric fields are combined is shown in
Fig.~\ref{shortp}. The most frequent interval is 0.25 days,
corresponding to one spin period between two successive
observations in the medium band photometer and accounts for 50
percent of the intervals. The origin of the other intervals can be
read in Table~\ref{tab:interval}, by adding in some instances one
or two revolutions of 0.25 days.

\begin{table}
\setlength{\tabcolsep}{2.5mm}
\begin{center}
 \caption{ Frequency distribution of the number of observations in
continuous sequences. In the 2nd and 3rd columns only transits in
the astrometric FOV are considered (that is to say in the BBP
filters) while the last column includes also the MBP filters. The
table gives the frequency of sequences of different lengths,
according as the first field in the sequence is the preceding or
following astrometric field of view (col 2 \& 3), while for the
MBP, due to the size and position of the detectors, all sequences
start with few consecutive observations in MBP, followed by
transits including only  BBP and MBP, before additional crossings
of the MBP.}\vspace{5mm}
  \begin{tabular}{lcc|r}
   \hline &&& \\[-4pt]
& \multicolumn{2}{c|}{Sequence starts with }\\[3mm]
      \cent{length} &   \cent{\pfov} &  \mc{1}{c|}{\ffov} &  \cent{{\sc MBP}} \\[3mm]
    \hline &&&\\[-4pt]
 1      & 0.27    & 0.63  & 0.00   \\[1mm]
 2      & 0.60    & 0.20  & 0.00   \\[1mm]
 3      & 0.05    & 0.11  & 0.06   \\[1mm]
 4      & 0.04    & 0.02  & 0.13   \\[1mm]
 5      & 0.01    & 0.02  & 0.33   \\[1mm]
 6      & 0.01    & 0.00  & 0.22   \\[1mm]
 7      & 0.00    & 0.00  & 0.07   \\[1mm]
$>7$    & 0.01    & 0.01  & 0.19   \\[2mm]
  \hline
  \end{tabular}
\label{tab:sequences}
\end{center}
\end{table}

\subsection{Long term succession}

The short term succession within an epoch is essentially driven by
the relative position of the fields and the spin period, which
control the time needed to rotate from one field to the next. The
transverse motion which leads to the end of a sequence and later
to the beginning of a new one is due to the slight displacement of
the spin axis on the precession cone of $\sim$ 4 degrees per day.
Therefore the return of an epoch of observation is primarily
determined by the precession motion and the yearly solar motion on
the ecliptic.

We have seen in the previous section, that the average length of a
sequence in the astrometric fields only, is very close to two.
With a total of $\sim$ 80 observations in total
(Fig.~\ref{fig:nmesbeta}), this gives about 40 epochs distributed
over the 1800 days of the mission, or a typical interval between
epochs of 45 days. If we take the combined observations in the two
photometers instead, we have 280 observations and an average
sequence of 6.5 observations, that is to say a return again every
45 days. However we have seen that the total number of
observations per star is also  function of the ecliptic latitude,
as any other effect tied to the precessional motion. The actual
number as a function of the latitude can only be obtained through
a simulation with a fairly large number of stars.

\begin{figure}
  \centerline{\includegraphics[width=7cm]{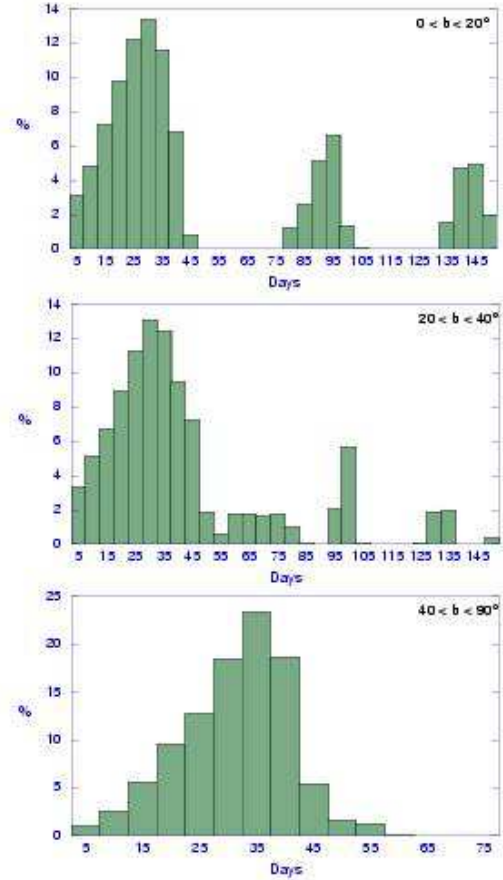}}
  \caption{Frequency distribution (as a function of the ecliptic
    latitude) of the intervals of time between epochs of observations
    in the three photometric fields of view. The recurrence is
    primarily driven by a combination of the precessional motion of
    \gaia\ spin axis and the yearly solar motion.}\label{longp}
\end{figure}

The frequency distribution is shown in Fig.~\ref{longp} for the
ecliptic region, the intermediate latitude and the extended polar
cap. These distributions are very informative, and indeed more
relevant for the purpose of variable analysis than a weakly
scattered distribution about the average interval would be. Below
$\beta2 =40 \degr$, there is quite an extensive coverage between 5
to 150 days, with two holes in the ranges 50--80 days and 100--130
days. At higher latitude, the scatter is less pronounced, with a
smaller period between two epochs, meaning more observations.
Clearly the period recovery of variable stars will depend, not
strongly indeed, on the ecliptic latitude just because of the time
sampling. The evaluation of this feature is the main object of the
coming sections of this paper. A similar conclusion applies to the
periods of spectroscopic binaries determined from the radial
velocity measurements as found by Pourbaix \& Jancart (2003).

\section{Simulations}
\subsection{The time sampling}
As said earlier a catalogue of 10\,000 directions uniformly
distributed has been produced. From the position and size of the
different fields of view and the scanning law it was possible to
sample the 1800 days of the mission and determine at any time which
regions of the sky are observable by each instrument. When a star lies
in a field, one computes then accurately the date of the crossing of
the central line of the field, together with its ordinate and
transverse velocity and the index of the field. Many refinements are
used to speed up the runtime by eliminating quickly many of the stars
not observable within $\sim$ a day. The output is finally re-organised
to produce chronological time series for each of the 10\,000 stars.
This gives at the end a set of about 3 million observations scattered
over 5 years. This file of realistic observing sequences is then the
basis of the photometric simulation and the variability analysis.

\begin{figure}
  \centering%
  \mbox{\psfig{file=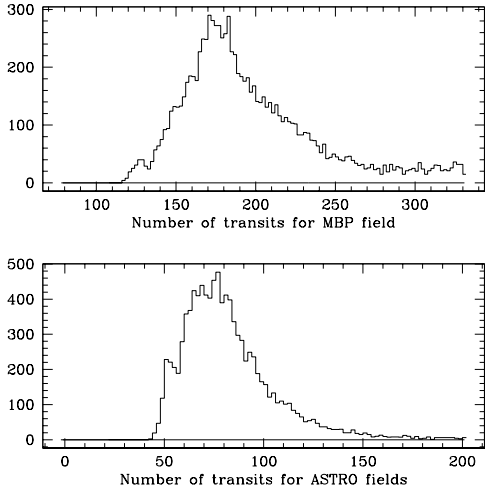,width=70mm}}%
  \caption{\label{fig:histo_obs}
    Histograms of the number of measurements for the
    astro-fields (both following and preceding) and for the MBP field.
    The extended scatter towards the largest values comes from the
    observations around $40 \degr$ of ecliptic latitude.  }
\end{figure}

The distribution of the number of individual observations (an
observation at a particular date corresponds to the crossing of
one field of view) is shown in Fig.~\ref{fig:nmesbeta} as a
function of the ecliptic latitude for the $10\,000$ sources of the
simulation. There is a strong dependence resulting from the
scanning law, with mid-latitude region over observed compared to
the mean, while the ecliptic region is observed less frequently
than the average. The same pattern is common to the main
astrometric field, which includes the BBP filters, and to the MBP
field. However due to the larger across-scan extension of the
latter, about twice as many observations are made in the latter field.
This appears clearly in Fig.~\ref{fig:histo_obs} which displays
the frequency distribution of the number of observations in the
same two fields of view. The scatter in each field is significant
with variation from 40 to 150 in the astrometric fields and 100 to
300 in the MBP field.

\begin{figure}
  \centering%
  \mbox{\psfig{file=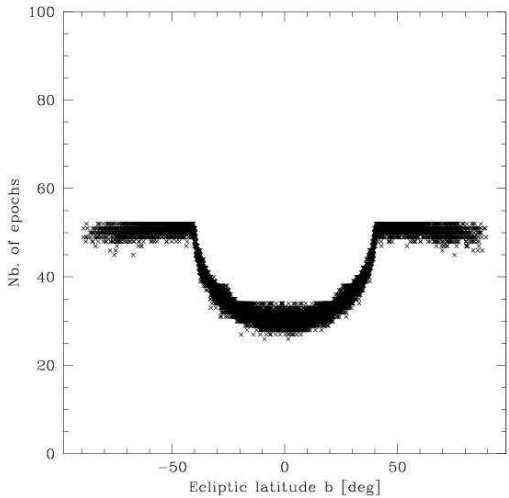,height=70mm,width=70mm}}%
  \caption{\label{fig:epoch}
    Number of clustered groups as function of ecliptic latitude $
    \beta2 $. The pattern is much simpler than that for the total
    number of transits (Fig.~\ref{fig:nmesbeta}), with only two
    regimes : low and high ecliptic latitude.}
\end{figure}

Quite as important as the number of observations is the number of
different epochs at which the signal is sampled.  The observations
are clustered into small groups of $\sim $ 2 to 10 observations
within a day, before a new cluster shows itself after several
weeks. The number of epochs based on the same simulation plotted
in Fig.~\ref{fig:epoch} shows that there are essentially two
regimes: the ecliptic zone with $\pm 30 \degr$ with about 30
observational epochs over the five years and the region of upper
ecliptic latitude, above $ 40 \degr$, with $\sim 50$ different
epochs. This feature impacts  significantly on the variability
analysis probably more than the total number of observations,
since it is directly related to the ability to recover the
characteristics of periodic phenomena. Even after a normalization
over the $S/N$ values, there should remain a systematic effect
with the ecliptic latitude.

\subsection{The photometric signal}
For each star one generates the photometric variability, modelled
by a simple sinusoidal signal with one frequency. More complex
light curves have been tested with several harmonics representing
non-sinusoidal oscillations of pulsating variables or eclipsing
binaries without affecting the main conclusions of this study,
although the case of the eclipsing binaries is the source of specific
problems. Therefore our baseline simulated signal has the form,
$$
  s(t)=  S_0 + A \sin(2 \pi \nu t + \phi) + \mbox{noise}
$$

where $A$ is the amplitude of the variability, $\nu$ the
frequency, $\phi$ the phase. One should note that variable stars
are usually characterised by the amplitude of variability, that is
to say by the peak to peak amplitude $2*A$. The noise added to the
deterministic signal is a Gaussian random variable of zero mean
with a standard deviation determined by the signal-to-noise ratio
of each experiment. Although we have also experimented with
multi-periodic signals, the results presented here refer to purely
periodic variables.

All the simulations are scaled by the $S/N$ ratio (defined here as
$A/$noise). This is a convenient parameter which makes our results
independent of the star magnitude. However as a result of the
systematic variation of the number of observations with the ecliptic
latitude, the dominant effect in the variability analysis beyond the
$S/N$ ratio, is the ecliptic latitude. In order to avoid this bias, we
have used a normalised definition of the signal-to-noise ratio, by
dividing the single observation noise in $S/N$ by $n^{1/2}$. More
precisely we have,
\begin{equation}\label{norm_sn}
  \chi = S/N \times \sqrt{n/80}
\end{equation}
so that $\chi \sim 1$ for an average observation with $S/N =1$,
while at $\beta2 \sim 40 \degr$, $ \chi = 1$ is equivalent to $S/N
\sim 0.65$. Therefore any remaining effect with the ecliptic
latitude will arise more from the time sampling than from the
systematic difference in the number of observations.

Simulated variables are generated for a grid of frequencies and
$\chi$. For each point of this grid, and for all the sources a
period search is performed on the signal, without any {\sl a
priori} information, using primarily an optimised version of the
Lomb-Scargle algorithm \cite{press92} or a frequency analysis
based on orthogonal decomposition developed by one of us (FM). The
rate of success in the period retrieval for every grid point is
the main output of the program.  Regarding the size of the
computation, we have sampled about 10 periods (from the slowest
Miras to the shortest pulsating variable with periods of few
hours) and between 3 and 5 values of the normalised $S/N$. Each of
the 10\,000 stars (or 5000 for the runs limited to one hemisphere)
has been used 5 times, with independent random noise on the
observations. This gives in total 50\,000 period searches for
every grid point.  In total the number of computed "periodograms"
reaches nearly 1.4 million.


\section{Results}
\begin{figure}
  \centering%
  \mbox{\psfig{file=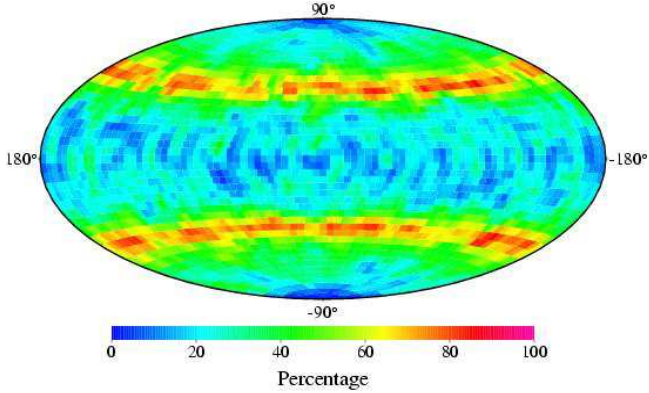,width=85mm,clip=true}}%
  \caption{\label{fig:aitoff}
    Rate of correct detections for $S/N=0.75$ and a single period of 0.2 days,
    using only the astrometric fields.  We are here in a range of
    small signal-to-noise ratio.  For certain regions, the rate reaches
    nearly 0\% and for other regions where the number of measurements is
    high, it reaches nearly 100\%.  The small effects in longitude follow from
    the non-uniform coverage observation density as of function of longitude over 5 years.}
\end{figure}
\subsection{Rate of detection}
Results consist primarily of correct rate of detection as a function
of the period, the signal-to-noise ratio and the position on the sky,
this as a result of the inhomogeneous time sampling of \gaia.
Fig.~\ref{fig:aitoff} gives for a single grid point ($ P = 0.2$ day
and the unnormalised $S/N = 0.75$) the sky distribution of the correct
rate of period retrieval. The main pattern reproduces the scatter of
the number of observations with the ecliptic latitude, which has been
subsequently eliminated with the normalised $S/N$.  In the range of
latitude around $40 \degr$ the rate is close to $80$~percent, and
between $ 20$ and $50$~percent outside this area. The rate falls off
very quickly with lower signal-to-noise ratio, but on the other hand
grows quickly to $100$~percent for $S/N > 1.5$.

\begin{figure}
  \centering%
  \mbox{\psfig{file=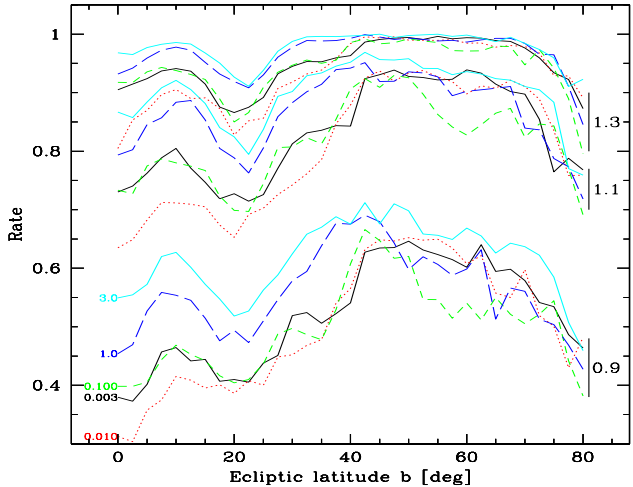,height=70mm,width=90mm}}%
  \caption{\label{fig:taux}
    Rate of correct detection for normalised signal-to-noise ratios
    $\chi$ (written on the right) and frequencies [cycle/day] (written
    on the left for the normalised signal-to-noise ratio of 0.9, the same
    colour/gray scale and line type are applied for the higher signal-to-noise ratios).  }
\end{figure}
Since there are only small secondary modulations with the ecliptic
longitude, one can perform one dimensional analyses by looking at the
dependence with the ecliptic latitude for various grid-points.
Results are shown in Fig.~\ref{fig:taux}. There are three sets of
curves for $\chi = 0.9, 1.1, 1.3$ and different periods (8h, 1, 10,
100 and 330 days).
\begin{itemize}
\item It is clear, that despite the normalisation of the $S/N$, the
  effect with the ecliptic latitude has not fully vanished, indicating
  an influence of the number of epochs (the number of clusters of
  observations separated by few weeks shown in Fig.~\ref{fig:epoch}).
  There is a systematic loss in the period recovery around $\beta2
  \simeq 20 \degr$, not easily understood at the moment.
\item The rate of correct recovery grows quickly to $ 100 $ percent,
  with normalised $S/N \geq 1.5$.
\item Although the scatter is large, it seems that the shorter periods
  ($< 1$ day) are better recovered than the larger periods. However
  the difference is not conspicuous and this comes as a surprising
  (and welcome) result, as it was thought that very short periods
  would be very difficult to recover.
 \end{itemize}

\subsection{Aliasing and false periods}
\begin{figure}
  \centering%
  \mbox{\psfig{file=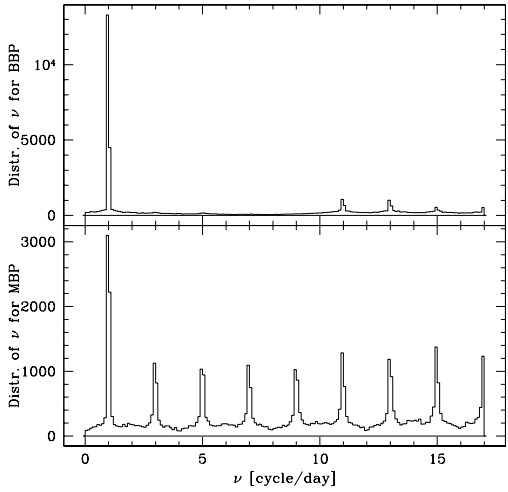,height=70mm,width=70mm}}%
  \caption{\label{fig:termfreq}
    Example of terminal frequencies for an initial signal of
    $S/N=0.75$ and frequency $\nu=1.000$. It reflects the pattern of
    individual periodogram, which bears the signature of the spectral
    window.}
\end{figure}

With irregular sampling, and within the range of periods analysed
here, one should not expect too much aliasing \cite{eyer99},
since the somewhat randomness of the sampling limits the building
up of coherent signature at regularly spaced frequencies. However,
at low signal-to-noise levels, the recovery rate is low, and the
search program may end up in a normal exit with no periodic signal
found (the favourable outcome) or with the identification of a
period (possibly wrong), without warning. It is then of interest
to look at what kind of periods are found as a function of the
true period. This is illustrated in Fig.~\ref{fig:termfreq} for
the BBP (same field as the astrometric detector) (top) and the MBP
fields (bottom). The simulated variable has a period of one day
and $S/N = 0.75$ in this example. In the BBP, when the period is
not recovered (about half of the cases tested) the final erroneous
frequency can be almost anything, with a higher probability to
frequencies of 11 and 13 cycles per day, coming from the
convolution with the spectrum of the observing window. So, unless
further assumptions, there is no way to decide from the output of
the period search whether the period found is correct, although a
significance level is easily produced and could be used as a
warning, but not as a measurement of the uncertainty of the
period. In the case of the MBP, aliased results are clearly
visible, regularly spaced with $\delta \nu = 2$ cycles/day,
corresponding to half of the satellite spin frequency (period of 6
hours). This comes from successive observations regularly sampled
every revolution within a cluster of observations, responsible for
the aliasing and leading to a wrong determination of the period.
This again cannot be avoided with the present design of the
instrument, but fortunately limited to those observations with low
SNR.

\section{Conclusion}

In this paper we have investigated the retrieval of periods of
variable stars from observations to be carried out during the \gaia\
mission. We have found, that with adequate frequency analysis
algorithms, one can recover period of regular variable stars even with
signal-to-noise ratio around unity. It is also shown that the very
irregular time sampling expected from \gaia\ allows to determine
frequencies much higher than the upper frequency estimated from a
straightforward evaluation of the Nyquist boundary (cf. Eyer \&
Bartholdi 1999), and with virtually no aliasing. These two features
are very good news for the preparation, and subsequent scientific
exploitation, of the mission.

However, from this investigation we have noted that the frequency
analysis without initial assumption on the frequency range may be very
demanding in term of computer resources, hardly compatible with the
goal of analysing from scratch $\sim 10^8$ sources. Therefore a major
effort is needed during this preparatory phase to develop optimised
software, to estimate quickly and efficiently possible periods or
ranges of periods, before a thorough and accurate search can be
launched on a well bounded frequency range. An estimate of the level
of significance of each peak height should be also a built-in feature,
alongside the possibility of searching for harmonics or a second
independent period.

\section{Acknowledgements}
We would like to thank Michel Grenon for interesting discussions
and Floor van Leeuwen for his valuable comments on the article.

\section*{Appendix : Derivation of equation 1}

Let a star  observed in one field and crossing the field in the
along-scan direction at the ordinate $y$. The field height is $h$
and $ 0<y<h$ during the transit. This crossing may happen at any
ordinate with a uniform probability, so that $P(\bf{Y} < y) =
y/h$. After an interval of time $\Delta t$ (one of the possible
intervals between the preceding and following fields listed in
Table~\ref{tab:interval}) the rotation of the satellite has
brought the line of sight of the second field in the same
direction  and the star may be observed again, provided its
transverse motion during $\Delta t$ has not been too large.
Mathematically the star cannot be observed again if $ y +V_t\
\Delta t > h$, meaning that its new ordinate is larger than the
field height (we have restricted to the case $V_t
> 0$ to simplify the discussion, but this does not alter the final
result for symmetry reasons). Therefore the probability that the
star is not observed in the second field when the transverse
velocity is $V_t$ is,

\begin{eqnarray*}
\lefteqn{ P(({\bf{X=1}})|V_t) = }\\ \\
   & &  \left\{\begin{array}{lcl}
    1  & \mbox{\ if \ } & V_t\ \Delta t \ge h \\[4mm]
    P({\bf{Y}} > h-V_t\,\Delta t) =\frac{V_t\,\Delta t}{h}  & \mbox{\ if \ } &  V_t\ \Delta t \le h
     \end{array}
  \right.
\end{eqnarray*}
In the first case the transverse velocity is large enough for the
transverse motion during $\Delta t$ to be greater than the field
width. In the second case there exists a region in the initial
field of view where even after the time $\Delta t$ the ordinate of
the star is still less than the field width. The above
probabilities are conditioned on the transverse velocity $V_t$ and
one must now introduce its random distribution. Assuming that the
star can be anywhere on the scan circle  with a uniform
probability, one has $V_t = V_{\mbox{max}}\ \sin(\phi)$ where the
maximum velocity $V_{\mbox{max}} = 0.170$ deg/h and $\phi$ is
uniformly distributed between $0$ and $2\pi$. The final
probability is just the average of the conditional probability
over the distribution of the transverse velocity (restricted here
to the positive values). The case when  $V_{\mbox{max}}\ \Delta
t/h  \le 1$ is very simple with
$$\label{firstcase}
    P({\bf{X=1}}) = \frac{2}{\pi}\, \frac{V_{\mbox{max}}\ \Delta
    t}{h}\ \int_0^{\pi/2} \, \sin \phi\, d\phi =\frac{2}{\pi}\, \frac{V_{\mbox{max}}\ \Delta
    t}{h}
$$
In the second case the integration extends up to $\phi_m$ such
that  $V_{\mbox{max}}\ \Delta t \sin\phi_m = h$, or $ \phi_m =
\arcsin (h/V_{\mbox{max}}\ \Delta t )$, and beyond this angle the
conditional probability is one. This yields for $V_{\mbox{max}}\
\Delta t/h  \ge 1$,
$$\label{firstcase}
    P({\bf{X=1}}) = \frac{2}{\pi}\, \frac{V_{\mbox{max}}\ \Delta
    t}{h}\ \int_0^{\phi_m} \, \sin \phi\, d\phi + \frac{2}{\pi}\ \int_{\phi_m}^{\pi/2} \, d\phi
$$
giving with $u = h/(V_{\mbox{max}}\ \Delta
    t)$ (hence $u < 1$ here),
$$
P({\bf{X=1}}) =  \frac{2}{\pi}\,\frac{1}{u}\left[ 1 - \left(1-
u^2\right)^{1/2}\right]+ \left[ 1- \frac{2}{\pi}\, \arcsin u
\right]
$$
which is the same as Eq.~\ref{probability}.
\end{document}

%% file: my_def_aa.txt
\def \gaia {{Gaia}}
\def \pfov {{\sc pfov}}
\def \ffov {{\sc ffov}}

\def \hip {{\sc hipparcos}}

\def \ccd  {{\sc ccd}}
\def \ccds  {{\sc ccd}s}

\def \etal {{\sl et al.}}
\def\cent{\multicolumn{1}{c}}
\def\mc{\multicolumn}

\newcommand{\fov}{{\sc fov}}

%% file: ME667rv2_title.bbl
\begin{thebibliography}{}
\bibitem[Eyer \& Bartholdi, 1999]{eyer99}
     Eyer, L., \& Bartholdi, P. 1999, A\&AS, 135, 1
\bibitem[Eyer \& Cuypers, 2000]{eyer00}
     Eyer, L., Cuypers, J.
     2000,
     in ASP Conf. Ser. 203,
     The Impact of Large-Scale Surveys on Pulsating Star Research,
     ed. L. Szabados, \& D. W. Kurtz, Proc IAU Colloq. 176, 71
\bibitem[Eyer \& Grenon, 1997]{eyer97}
     Eyer, L., Grenon, M. 1997, ESA SP-402, 467
\bibitem[Jordi, 2003]{jordi03}
     Jordi, C., \etal\ 2003, in
     \gaia\ Spectroscopy, Science and Technology, 209--220,
     ASP Conf. Ser. 298,
     ed. U. Munari
\bibitem[Mignard, 2001]{mignard01}
     Mignard, F. 2001, A practical scanning law for \gaia\
     simulations, Technical Note GAIA\_FM\_010\
\bibitem[Perryman \etal, 2001]{perryman01}
     Perryman, M.A.C., de Boer, K.S., Gilmore, G., \etal\
     2001,
     A\&A, 369, 339
\bibitem[Pourbaix \& Jancart 2003]{pourbaix03}
     Pourbaix, D., \& Jancart, S. 2003,
     in ASP Conf. Ser. 298,
     Gaia Spectroscopy, Science and Technology,
     ed. U. Munari,
     345
\bibitem[Press et al., 1992]{press92}
     Press, W. H., Teukolsky, S. A., Vetterling, W. T.,
     Flannery, B. P. 1992, Numerical Recipes in Fortran.
     Cambridge University Press
\end{thebibliography}
